\documentclass{elsarticle}

\usepackage{hyperref}
\usepackage{xcolor}
\usepackage{selinput}
\usepackage{ulem}
\journal{Physica B}

\begin{document}

\begin{frontmatter}

\title{Crystal Growth of new charge-transfer salts based on $\pi$-conjugated donor molecules}

\author{Antonia Morherr*\fnref{myfootnote1}\corref{mycorrespondingauthor}}
\ead{morherr@stud.uni-frankfurt.de}
\author{Sebastian Witt\fnref{myfootnote}}
\author{Alisa Chernenkaya\fnref{sec,Mainz}}
\author{Jan-Peter B\"acker\fnref{myfootnote}}
\author{Gerd Sch\"onhense\fnref{Mainz}}
\author{Michael Bolte\fnref{Chemie}}

\author{Cornelius Krellner\fnref{myfootnote}}

%\fntext[mymainaddress]{Max-von-Laue-Straße 1, 60438 Frankfurt am Main, Germany}

\fntext[myfootnote]{Physikalisches Institut, Goethe-Universit\"at Frankfurt am Main, 60438 Frankfurt am Main}
%\fntext[mycorrespondingauthor]{Max-von-Laue-Straße 1, 60438 Frankfurt am Main, Germany}
\fntext[sec]{Graduate School Materials Science in Mainz, 55128 Mainz}
\fntext[Mainz]{Institut f\"ur Physik, Johannes Gutenberg-Universit\"at, 55099 Mainz}
\fntext[Chemie]{Institut f\"ur anorganische Chemie, Goethe-Universit\"at Frankfurt am Main, 60438 Frankfurt am Main}

%\cortext[mycorrespondingauthor]{Max-von-Laue-Straße 1, 60438 Frankfurt am Main, Germany}

%\address[mymainaddress]{Max-von-Laue-Strasse 1, 60438 Frankfurt am Main, Germany}

\begin{abstract}
New charge transfer crystals of $\pi$-conjugated, aromatic molecules (phenanthrene and picene) as donors were obtained by physical vapor transport. The melting behavior, optimization of crystal growth and the crystal structure is reported for charge transfer salts with (fluorinated) tetracyanoquinodimethane (TCNQ-F$_x$, x=0, 2, 4), which was used as acceptor material. The crystal structures were determined by single-crystal X-ray diffraction. Growth conditions for different vapor pressures in closed ampules were applied and the effect of these starting conditions for crystal size and quality is reported. 
The process of charge transfer was investigated by geometrical analysis of the crystal structure and by infrared spectroscopy on single crystals. With these three different acceptor strengths and the two sets of donor materials, it is possible to investigate the distribution of the charge transfer systematically. This helps to understand the charge transfer process in this class of materials with $\pi$-conjugated donor molecules.

\end{abstract}

\begin{keyword}
charge transfer, TCNQ, physical vapor transport
\end{keyword}

\end{frontmatter}

%\linenumbers

\section{Introduction}

Designing new organic materials and tuning their physical properties by adding or removing charges has a long history\cite{Bechgaard1981}, but the detailed understanding of mechanisms is still a matter of current research. Organic charge-transfer (CT) salts offer a huge variability in crystal structures and physical properties ranging from metallicity over superconductivity to Mott insulators \cite{Kist1974,Lang2007} on one side. Organic semiconductors on the other side find application in various diodes and organic electronics due to their flexibility and relatively high mobility \cite{Podz2003}. Also, a new class of multiferroicity was found in organic charge transfer salts and introduced a new research field \cite{Lunk2012}.

In the last years, the class of phenacenes attracted a lot of attention. For potassium doped phenanthrene (n=3 benzene rings) \cite{Wang2011}, picene (n=5) \cite{Mitsu2010} and dibenzopentacene (n=7) \cite{Dbpen2012} superconducting phases were discovered with a T$_C$ of 5$\,$K (n=3), 18$\,$K (n=5) and 33$\,$K (n=7), respectively. The charge transfer process in crystals based on these three compounds is of interest, as the reproducibility of these samples is still under debate \cite{Heguri2015} and only few experimental studies have been reported. 
In this paper, we expand the sets of charge transfer (CT) salts based on aromatic donor molecules as reported for phenanthrene/TCNQ \cite{Dobr2014} and picene/TCNQ-F$_4$ \cite{Mahns2014} and present the crystal structures of new charge transfer salts phenanthrene/TCNQ-F$_x$ (x=2, 4) and picene/TCNQ-F$_x$ (x=0, 2). 
With respect to the weak donor ability of phenanthrene and picene (HOMO: 6.1$\,$eV, 5.5$\,$eV \cite{HOMOpic}) and the strong acceptors TCNQ, TCNQ-F$_2$ and TCNQ-F$_4$ (LUMO: 4.2$\,$eV, 4.55$\,$eV and 5.24$\,$eV\cite{HOMOLUMO}) neutral CT are expected. With the two known structures of PHN/TCNQ and PIC/TCNQ-F$_4$, we yield a set of two donors with three different acceptors. By changing the fluorine content in the acceptor, the overall charge transfer is increased. These novel materials expand the materials basis for the recently suggested HOMO-LUMO engineering \cite{HOMOLUMO} utilizing charge-transfer compounds.

\section{Experimental details}
\subsection{Crystal growth}
\begin{figure}[h]
\centering
  \includegraphics[width=9cm]{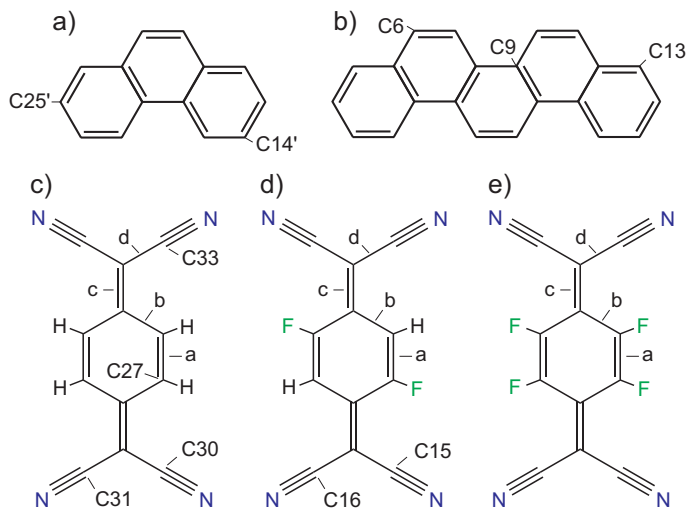}
  \caption{a) Phenanthrene and b) picene used as donor molecules and c) TCNQ, d) 2,5-TCNQ-F$_2$ and e) TCNQ-F$_4$ as acceptors are shown. The bond lengths a-d are marked in the acceptors to describe structural effects of charge-transfer. The colors representing different atoms in the molecules are as follows: grey (C), white (H), blue (N) and green (F). The numbers of C atoms refer to selected distances between acceptor and donor molecules, discussed within this work.}
  \label{fgr:DandAnumber}
\end{figure}

Phenanthrene (PHN) and picene (PCN) as donor molecules (D) were purchased by TCI with a purity of $>$97\% and $>$99\%, respectively. TCNQ, TCNQ-F$_2$ (2,5-configuration) and TCNQ-F$_4$ as acceptors (A) were also purchased by TCI with purities of $>$99\%, $>$98\% and $>$98\%, respectively. These starting molecules are shown in Figure \ref{fgr:DandAnumber}. The acceptor strength increases for higher fluorinated acceptors with an electron affinity of 2.85$\,$eV for TCNQ; 3.02$\,$eV for TCNQ-F$_2$ and 3.20$\,$eV for TCNQ-F$_4$ \cite{Emge1981}.

\begin{figure}[h]
\centering
  \includegraphics[width=8cm]{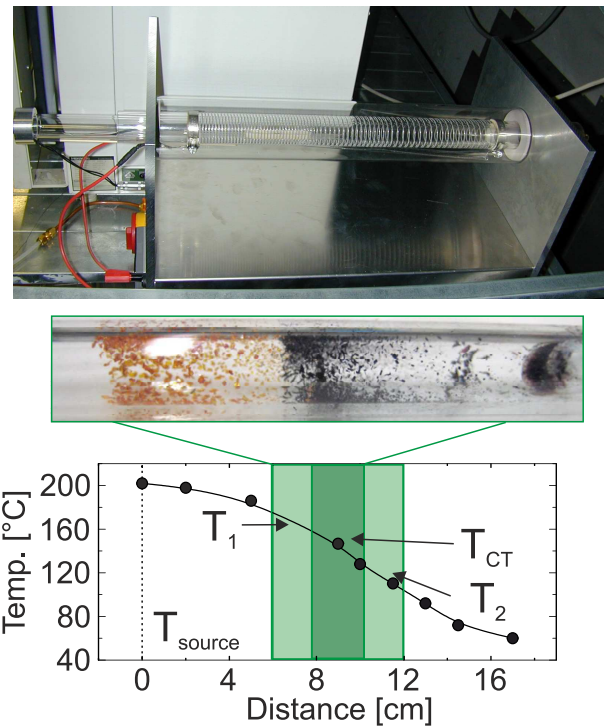}
  \caption{The PVT oven for crystal growth is shown on the top designed after the working principle used in Ref.~\citenum{Goldmann2004}. A typical temperature profile for phenanthrene based CT crystals is shown at the bottom with the hottest zone of around 200$^\circ$C. TCNQ resublimes first at T$_1$=170$^\circ$C as orange crystals, the CT complex (black) at T$_{CT}$=140$^\circ$C and the phenanthrene at T$_2$=110$^\circ$C as transparent crystals.}
  \label{fgr:Ofen}
\end{figure}

Single crystals were grown by physical vapor transport (PVT)\cite{Kloc1997,Laud1998} in closed glass ampules. For this purpose, the starting materials (as powders) of D and A were ground together and transferred into a glass ampule, which was cleaned by ethanol and acetone before and baked in a box furnace for at least 24 hours to remove water (as described in Ref.~\citenum{Mahns2014}). The glass ampule was evacuated to $p=10^{-3}\,$mbar before closing. The experimental oven setup and a typical temperature gradient used in the experiments is shown in Figure \ref{fgr:Ofen} with T$_{source}$ as temperature at the hottest position with the source material and T$^x_{crystal}$ with x = 1, 2, CT, where \textit{x} denotes the crystal growth temperature of material 1, material 2 and the CT complex, respectively. Typical crystal growth conditions are listed in table \ref{tbl:growth}. 

\begin{table}
\small
  \caption{\ Growth parameters for phenanthrene and picene based charge-transfer salts.}
  \label{tbl:growth}
  \begin{tabular}{llllll}
    \hline
          &   m$_{donor}$ [mg]  &  m$_{acc.}$ [mg]  &  $T_{max}$ [$^\circ$C]  & $T_{CT}$ [$^\circ$C] & stoichiometry \\
    \hline
    PHN/TCNQ             &  26.8 & 13.2 & 200 & 160 - 130 & 1:1\\
    PHN/TCNQ-F$_2$ & 17.2 &  2.6  & 200 &  180 -160 & 1:1\\
    PHN/TCNQ-F$_4$ & 21.8 & 12.1 & 200 & 186 - 147 & 1:1\\
    \hline
    PCN/TCNQ             & 10.8 & 15.9 & 240 & ca. 200 & 1:1\\
    PCN/TCNQ-F$_2$  &  26.4 & 18.2 & 245 & 220 - 185 & 1:1\\
    PCN/TCNQ-F$_4$  & 17.1 & 20.1 & 240 & ca. 200 & 1:1\\
    \hline
  \end{tabular}
\end{table}

The crystal growth zone lies in a region with a steep temperature gradient, which results in the best crystal growth as mentioned and suggested in Ref.~\citenum{Rosenberger}. The growth procedure lasted ca.$\,$50 hours until all starting material is sublimed and subsequently the ampule was removed from the hot oven. The section with the crystal growth zone of an ampule after the growth procedure of phenanthrene/TCNQ is also shown in Figure \ref{fgr:Ofen}. Here, TCNQ (orange) resublimes at higher temperatures (T$_1$) compared to the (black) CT complex (T$_{CT}$) and (transparent) phenanthrene (T$_2$). All charge transfer crystals grow in a 1:1 stoichiometry with the applied starting conditions. There is no other stoichiometry detected as for example 2:1 in Coronene$_2$/TCNQ-F$_4$ \cite{Yoshida2015}. Besides the growth of CT salts in a closed system, the oven can be used also for crystal growth or purification of organic semiconductors (molecular crystals with one type of molecule). Here, the crystal growth is performed under a stream of argon and a flat temperature gradient in the oven is necessary.      

\begin{figure}[h]
\centering
\includegraphics[width=9cm]{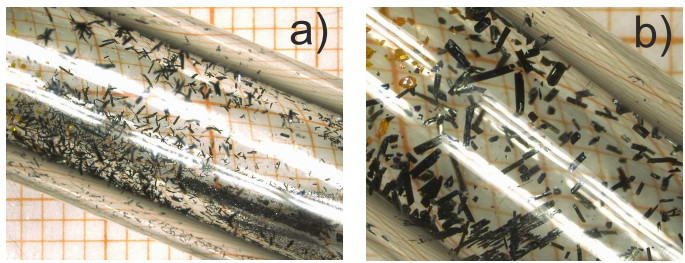}
\caption{Ampules with phenanthrene/TCNQ after growth in the charge transfer region for different pressures of argon a) $p=10^{-3}\,$mbar, b) $p=0.5\,$bar and c) 1$\,$bar.}
\label{fgr:PHNpressure}
\end{figure}

For phenanthrene/TCNQ small needle-like crystals were obtained with a source temperature $T_{source}$=200$^\circ$C. For phenanthrene/TCNQ-F$_4$, this method with a pressure of $p=10^{-3}\,$mbar results only in micro-crystallites (smaller than 50$\,\mu$m), where a separation from the glass walls after the growth procedure is not possible. For optimization of the crystal growth conditions, we varied the argon gas pressure inside the ampule. For this purpose, the ampule is first evacuated to $p=10^{-3}\,$mbar and than refilled with argon as inert gas. The pressure amounts to the value at room temperature. In Figure \ref{fgr:PHNpressure} typical results are shown for two different argon pressures for phenanthrene/TCNQ. The single crystals in the first case for a pressure of $p=10^{-3}\,$mbar are smaller than 1 mm and have a width and thickness of 250$\,\mu$m in average.
For the second case with an inert gas pressure of 0.5$\,$bar, larger crystals up to a length of 2$\,$mm in average are detected.

The improved crystal growth conditions are reflected in more block-like crystals by higher inert gas pressure.
This can be due to increased sublimation temperatures and modified growth rates for different pressures mentioned in Ref.~\citenum{Rosenberger}. Higher pressures might lead to a suppressed nucleation rate due to a smaller oversaturation at the ampule walls in the crystal growth regime. Therefore, the crystals grow slowly and a block-like habitus appears. 

\subsection{DTA}
\begin{figure}[h]
\centering
\includegraphics[width=12cm]{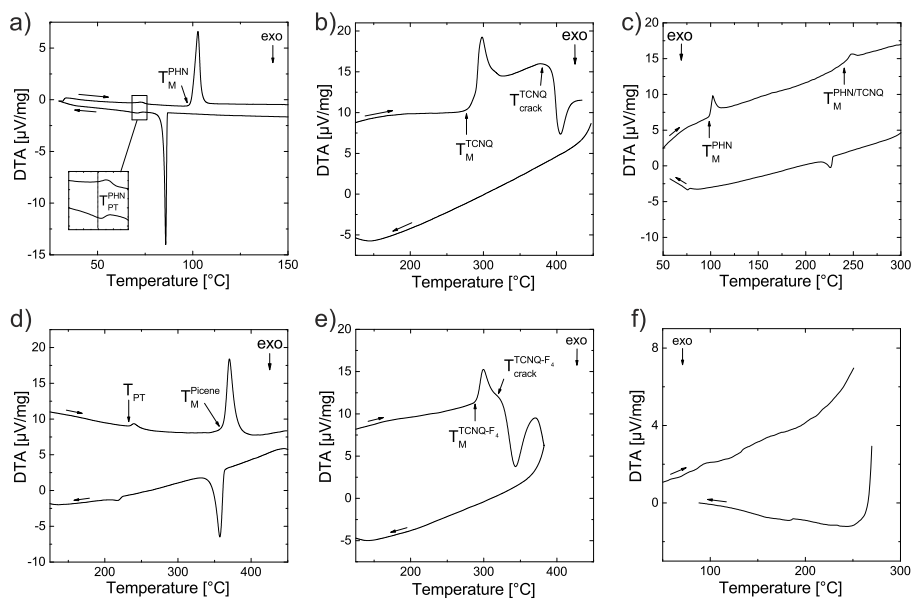}
\caption{DTA measurements for a) phenanthrene, b) TCNQ and c) both compounds in a closed DTA glass ampule are shown. In d)+e) the same measurements for picene and TCNQ-F$_4$ are shown. The peaks for melting points show up at $T_{M}^{PHN}$=100$^\circ$C, $T_{M}^{TCNQ}$=290$^\circ$C, $T_{M}^{PHN/TCNQ}$=242$^\circ$C, $T_{M}^{PCN}$=364$^\circ$C and $T_{M}^{TCNQ-F_4}$=290$^\circ$C. Phenanthrene shows a phase transition at $T_{PT}^{PHN}$=69$^\circ$C and a similar peak is measured for picene at $T_{PT}^{PCN}$=234$^\circ$C. The mixed compound of picene and TCNQ-F$_4$ in f) shows no CT peak as the cracking temperature of TCNQ-F$_4$ lies below the melting point of picene.}
\label{fgr:DTA}
\end{figure}

For a better understanding of the crystal growth conditions, differential thermal analysis (DTA) measurements were performed with phenanthrene, picene, TCNQ, TCNQ-F$_4$ and a mixture of donor and acceptor materials shown in Figure \ref{fgr:DTA}. The starting materials were filled into DTA glass ampules. The ampules were evacuated to $p=10^{-3}\,$mbar, refilled with 0.5$\,$bar argon and closed as described for the growth ampules above. The measurements were obtained by a STA 409 from NETZSCH and a DTA sample holder (type E). For heating and cooling, rates of 5$\,$K/min were applied. The four ampules with pure phenanthrene (0.16$\,$mmol), picene (0.089$\,$mmol), TCNQ (0.047$\,$mmol) and TCNQ-F$_4$ (0.041$\,$mmol) were heated to temperatures above the melting points of the different materials and then were cooled down again. Both processes were recorded with DTA and are presented in Figure \ref{fgr:DTA}. 
Melting points in DTA experiments are defined as the onset of the peak. The melting point for phenanthrene was reported to be at 100$^\circ$C \cite{Ueberreiter} and TCNQ melts around 293$^\circ$C \cite{Acker1962}. These temperatures are reproduced in DTA measurements shown in Figure \ref{fgr:DTA} for a) phenanthrene and b) TCNQ with onsets at T=100$^\circ$C and at T=290$^\circ$C, respectively. An additional peak appears for phenanthrene at 69$^\circ$C. This peak shows the phase transition (PT) to the high temperature crystal structure for phenanthrene as reported in Ref.~\citenum{Ueberreiter}. By cooling down phenanthrene, the two peaks were also detected with T$_{cryst.}=87^\circ$C and T$_{PT}$=73$^\circ$C for the low temperature crystal structure. 
For TCNQ (see Figure \ref{fgr:DTA} b)) an exothermal peak appears in the heating curve at T$_{crack}$=375$^\circ$C after the melting point at T$_{M}$=290$^\circ$C. The recrystallization peak is missing in the cooling part, which means, that the exothermal peak in the heating curve signalized a decomposing of the TCNQ molecules. 
The ampule with both starting materials with a surplus of phenanthrene (0.037$\,$mmol of PHN and 0.024$\,$mmol of TCNQ) is first heated to a temperature of T=305$^\circ$C to be above the melting point of TCNQ, but far below the cracking temperature. Phenanthrene melts and starts to solve TCNQ above 100$^\circ$C. After this first run, the powder was dark red, confirming a charge transfer process. The ampule was heated again to 300$^\circ$C and cooled down with measuring the DTA signal.  
In the heating part of the curve, a peak around 100$^\circ$C appears again, which is due to the intact phenanthrene, not incorporated into the CT complex. The melting point of the CT appears at T$_{M}$=242$^\circ$C (see Figure \ref{fgr:DTA}c)). The peak also is observed in the cooling part at T$_{cryst.}$ = 228$^\circ$C meaning both compounds were intact and TCNQ was not decomposed in this temperature regime. 

The melting point for the complex phenanthrene/TCNQ lies in between the melting points for the single compounds, which looks reasonable and is in good agreement to the different crystal growth areas for acceptor, CT complex and donor\cite{Mahns2014,Mahns2015,Yoshida2015} and is also apparent in Figure \ref{fgr:Ofen}. Sublimation points could not be detected by DTA but are stronger depending on pressure than melting points. By using additional inert gas inside the ampule, the sublimation temperature is shifted to higher temperatures and sublimation rates are slowed down achieving larger crystals.\\
For picene (0.023$\,$mmol) and TCNQ-F$_4$ (0.023$\,$mmol), DTA measurements were also performed and presented in Figure \ref{fgr:DTA}d) together with the TCNQ data in Figure \ref{fgr:DTA}e). Also for picene a phase transition probably due to a structural transition is visible around 234$^\circ$C, similar to the one detected for phenanthrene.\\
For TCNQ-F$_4$ we detect an exothermal peak after the melting point, similar to TCNQ but at a lower temperature of 315$^\circ$C. The melting point of picene lies at 362$^\circ$C as shown in Figure \ref{fgr:DTA} and is therefore similar to the cracking temperature of TCNQ and above the cracking temperature of TCNQ-F$_4$. No detection of the melting point is possible for both complexes as the acceptor molecules will decompose before melting. 

The shown DTA measurements imply two important points for the improvement of crystal growth by PVT: To avoid cracking of the acceptor molecules, temperatures below 315$^\circ$C are necessary for crystal growth of pure TCNQ but also for CT salts based on these acceptors from the vapor phase. To avoid a much faster sublimation of phenanthrene at $p=10^{-3}\,$mbar resulting in a separation in crystal growth area and a small yield of charge transfer salts, additional argon pressure is essential. This increased pressure results in a much better crystal growth for all acceptor compounds with phenanthrene as donor molecule. 

\subsection{Single Crystal Structure}
Single crystal X-ray diffraction was performed on selected crystals to determine the crystal structure. 
The detailed information of crystal structures are summarized in the SI.

Data for phenanthrene/TCNQ-F$_2$, phenanthrene/TCNQ-F$_4$ and picene/TCNQ-F$_2$ were collected on a STOE IPDS II two-circle diffractometer with a Genix Microfocus tube with mirror optics using Mo\textit{K}$_\alpha$ radiation ($\lambda$ = 0.71073$\,$\AA{}) and were scaled using the frame scaling procedure in the \textit{X-AREA} program system \cite{Stoe2002}.
The structures were solved by direct methods using the program \textit{SHELXS} \cite{Sheldrick2008} and refined against \textit{F}$^2$ with full-matrix least-squares techniques using the program \textit{SHELXL} \cite{Sheldrick2008}.
Data for picene/TCNQ were collected on a Nonius KappaCCD four-circle diffractometer with graphite monochromated Mo\textit{K}$_\alpha$ radiation ($\lambda$ = 0.71073 \AA{}) and were processed with the DENZO/SCALEPACK software \cite{Otwinowski1997}. 
The structure was solved by direct methods using the program \textit{SHELXS} \cite{Sheldrick2008} and refined against \textit{F}$^2$ with full-matrix least-squares techniques using the program CRYSTALS \cite{Betteridge2003}.

The phenanthrene molecules in phenanthrene/TCNQ-F$_2$ and phenanthrene/TCNQ-F$_4$ are disordered about a centre of inversion over two equally occupied positions.
In picene/TCNQ-F$_2$, the F and H atoms bonded to the six-membered ring are disordered over two positions with a site occupation factor of 0.835(3) for the major occupied sites.

 \subsection{Spectroscopic analysis}
Infrared spectroscopy was measured on a small amount of needle-like single crystals with no preferred orientation of PHN/TCNQ-F$_x$ and picene/TCNQ-F$_x$ and these were compared to spectra of powder samples of the pure acceptors TCNQ, TCNQ-F$_2$ and TCNQ-F$_4$. The spectra were obtained by using a Nicolet 730 FT IR spectrometer between 4000 and 500$\,$cm$^{-1}$ at room temperature with a resolution of 1$\,$cm$^{-1}$.

\section{Results and discussion}

\subsection{Crystal structure of charge transfer salts}

\begin{figure}[h]
\centering
\includegraphics[width=9cm]{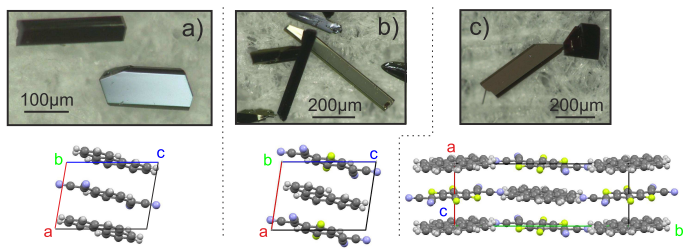} 
\caption{Single crystals and crystal structures of a) PHN/TCNQ; b) PHN/TCNQ-F$_2$ and c) PHN/TCNQ-F$_4$. The phenanthrene molecules are disordered in all structures.}
\label{fgr:PHNsalts}
\end{figure}

\begin{figure}[h]
\centering
\includegraphics[width=9cm]{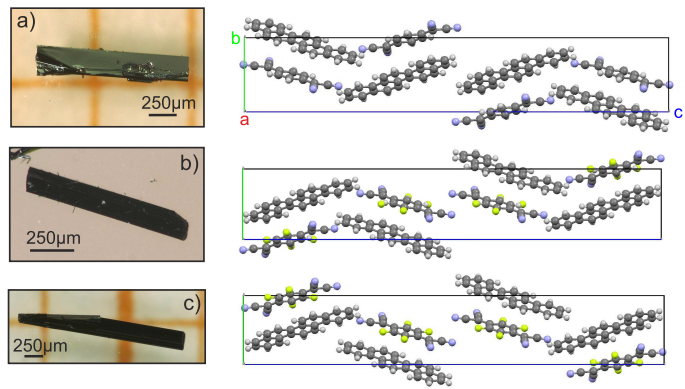}
\caption{Single crystals and crystal structures of a) PCN/TCNQ; b) PCN/TCNQ-F$_2$ and c) PCN/TCNQ-F$_4$. TCNQ-F$_2$ is disordered in the structure. The structures for these CT salts are isostructural.}
\label{fgr:PCNsalts}
\end{figure}

X-ray single crystal diffraction studies detect a ratio of 1:1 for donor and acceptor molecules and formation of mixed stack geometry in all compounds. Typical pictures of single crystals and the corresponding crystal structures resulting from single crystal X-ray diffraction are presented in Figure \ref{fgr:PHNsalts} for phenanthrene and in Figure \ref{fgr:PCNsalts} for picene based CT salts.

\begin{figure}[h]
\centering
\includegraphics[width=6cm]{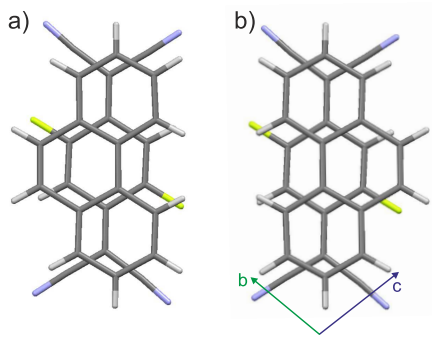}
\caption{Disordering for PHN/TCNQ-F$_2$ in a) left orientation and b) right orientation of the phenanthrene molecule.}
\label{fgr:PHNfehl}
\end{figure}

The crystal structure of phenanthrene/TCNQ-F$_2$ is isostructural to the reported for phenanthrene/TCNQ in Ref.~\citenum{Dobr2014}, whereas the space group (SG) changes from \textit{P}$\bar1$ to \textit{P}2$_1$/\textit{n} for phenanthrene/TCNQ-F$_4$.
Both, donor and acceptor molecules are planar and have similar dimensions. Especially, the longest axes  of phenanthrene and TCNQ(-F$_2$; -F$_4$) are roughly equal with 9$\,$\AA{} for phenanthrene and 8.4$\,$\AA{} for TCNQ (8.6$\,$\AA{} for TCNQ-F$_2$; 8.75$\,$\AA{} for TCNQ-F$_4$), respectively. The central part of TCNQ-F$_x$ and the benzene ring in the middle of PHN overlap with a little offset which is compensated by disorder, due to a mirroring of PHN on the long axis of the molecule shown in Figure \ref{fgr:PHNfehl} reported also for PHN/TCNQ\cite{Dobr2014}.

With a closer look to the stacking directions, the spacing becomes smaller for stronger acceptors as lattice parameter $a$ decreases from 6.719$\,$\AA{} in PHN/TCNQ over 6.682$\,$\AA{} in PHN/TCNQ-F$_2$ to 6.632$\,$\AA{} in PHN/TCNQ-F$_4$. 
By increasing the acceptor strength and geometry, the angle between donor and acceptor molecules and axis $c$ in \textit{P}$\bar1$ (axis $b$ in \textit{P}2$_1$/\textit{n}) becomes flatter. 
For picene/TCNQ and picene/TCNQ-F$_2$ the same space groups \textit{P}2$_1$/\textit{n} are revealed as for picene/TCNQ-F$_4$ reported in Ref.~\citenum{Mahns2014} and no disorder of the aromatic donor occurs in this set. Instead of the donor, the acceptor in the picene/TCNQ-F$_2$ structure is disordered. Here, no differentiation between one ordering of 2,5-configuration or mirroring at the bc-plane is possible. 
In picene/TCNQ-F$_x$ no space group changes are observed which can be due to different sizes of donor and acceptor in general (the longest axis is 13$\,$\AA{} for picene). This is compensated by packing in pairs of D and A in \textit{P}2$_1$/\textit{n} geometry as described in Ref.~\citenum{Mahns2014}.

\begin{figure}[h]
\centering
\includegraphics[width=9cm]{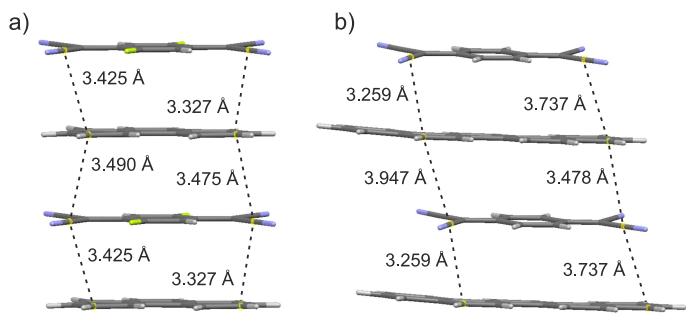}
\caption{Distances between D and A in a) phenanthrene/TCNQ-F$_2$ and b) picene/TCNQ charge transfer salts and bending of donor and acceptor molecules.}
\label{fgr:dimer}
\end{figure}

A structural indication for charge transfer in theses materials is, that shortest distances between e.g. atoms C16 and C14' in phenanthrene/TCNQ-F$_2$ or atoms C6 and C33 in picene/TCNQ are not equal as shown in Figure \ref{fgr:dimer} along the stacking axis. This effect occurs in all presented charge transfer salts but is the strongest for phenanthrene/TCNQ-F$_4$ with the largest bending of the cyano end groups. 
In general, the smallest C-C distance between donor and acceptor is in all cases below 3.5$\,$\AA{}  resembling a good $\pi-\pi$ overlap in all structures \cite{Schwoerer}.
An overview of the crystal structure data is shown in Table \ref{tbl:structuredata} and in the Supporting Information in more detail for each new CT complex. 

\subsection{Charge transfer by structural data analysis and infrared spectroscopy}

Another more pronounced property of CT salts is the amount of charge transfer going from donor to acceptor by formation of the CT compound. This value is relevant to classify CT salts and to understand their physical properties. One way to extract the CT of a system is to compare the bond lengths of the acceptor molecules. These changes of geometry inside the acceptor are a first indication to detect the amount of CT and were investigated for a wide range of TCNQ compounds \cite{Kist1982}. The same approach was also used for TCNQ-F$_2$ \cite{Emge1982} and TCNQ-F$_4$ \cite{Mahns2014}. In all cases, the pure acceptor compound has a quinoid structure. While charge transfer takes place, this quinoid structure in the acceptor changes to a more aromatic structure, which means that bond $c$ is extended and bond $b$ and $d$ are shortened to become a more aromatic character. This is shown in the literature for a variety of TCNQ salts. 
After Ref. \citenum{Kist1982}, we estimate the charge transfer by 

\begin{equation}
    \rho_{struct.}^{CT} = \frac{\alpha_{CT}-\alpha_{0}}{\alpha_{-1}-\alpha_{0}}
\end{equation}

where the parameter $\alpha$ is defined by the bond length ratio c/(b+d) (see Fig. \ref{fgr:DandAnumber}) and the indexes CT, 0 and -1 denote values for the CT compound, the pure acceptor and the acceptor anion, respectively. With this approach, the charge transfer was calculated for the presented compounds and shown in Table \ref{tbl:structuredata}.

\begin{table*}
\centering
 \small
  \caption{\ Structural parameters for the new charge transfer salts phenanthrene/TCNQ-F$_2$, phenanthrene/TCNQ-F$_4$, picene/TCNQ and picene/TCNQ-F$_2$ together with the known phenanthrene/TCNQ \cite{Dobr2014} and picene/TCNQ-F$_4$ \cite{Mahns2014}. The value of charge transfer obtained from structural changes in the acceptor molecule is shown in the last column.}
  \label{tbl:structuredata}
  \begin{tabular}{llllllllll}
    \hline
                & SG & $a$ [\AA{}]  & $b$ [\AA{}]  & $c$ [\AA{}]  & $\alpha$ [$^\circ$]  & $\beta$ [$^\circ$] & $\gamma$ [$^\circ$]  & C-C				$_{shortest}$ [\AA{}] & $\rho_{struct.}^{CT} [e^-]$\\
    \hline
    PHN/TCNQ             & \textit{P}$\bar{1}$       &  6.719     & 8.086         & 9.118       & 101.21      &  97.8     & 99.46        & 3.325  & -0.04\cite{Dobr2014}\\
    PHN/TCNQ-F$_2$  &   \textit{P}$\bar{1}$     &  6.682      & 8.078       &   9.277     &   101.66      &  96.31   &  100.27    & 3.327& 0.05\\
    PHN/TCNQ-F$_4$  &  \textit{P}2$_1$/\textit{n}       &   6.632      &  18.296     &    8.367    &    90          &   102.36   & 90          & 3.235 &  0.17  \\
    \hline
    PCN/TCNQ              &  \textit{P}2$_1$/\textit{n}       & 7.975       &  7.104       & 41.187     &  90             &  92.88   & 90            & 3.259 & 0.0\\
    PCN/TCNQ-F$_2$    & \textit{P}2$_1$/\textit{n}      & 7.905       & 7.081          & 42.177     & 90            &      91.64      &90       & 3.284 & 0.09 \\
    PCN/TCNQ-F$_4$      & \textit{P}2$_1$/\textit{n}     & 7.920        & 7.000      & 42.941      & 90             & 91.43       & 90          & 3.296 & 0.19 \cite{Mahns2014}\\ 
       
    \hline
  \end{tabular}
\end{table*}

Both sets of different acceptors result in a larger transfer for stronger acceptors, which is in good agreement with a smaller spacing between donor and acceptor in the structure and a larger overlap between orbitals of donor (D) and acceptor (A). 

\begin{figure}[h]
\centering
 \includegraphics[width=5cm]{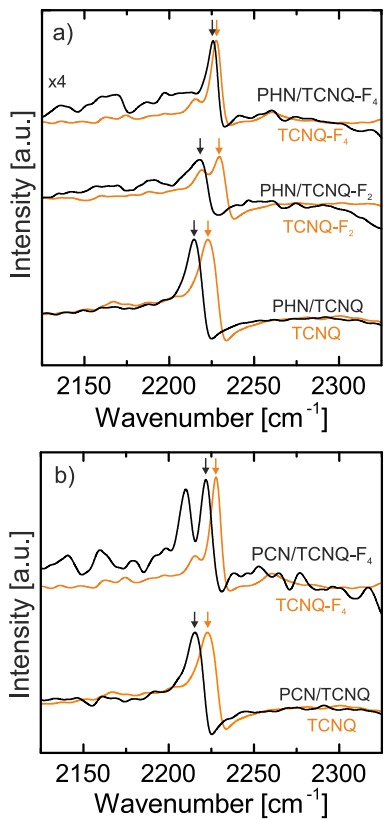}
 \caption{IR spectra of polycrystalline samples in the C$\equiv$N stretching mode regions for a) phenanthrene/TCNQ-F$_x$ (x=0, 2, 4)  and b) picene/TCNQ-F$_x$ (x=0, 4). The C$\equiv$N stretching is always red-shifted in the case of charge transfer compounds, compared to pure acceptors.}
 \label{fgr:IRCN}
\end{figure}

Another method to estimate the amount of charge transfer is the investigation with infrared spectroscopy \cite{Chapp1981,Mene1986}. This method was used widely for TCNQ and also in a few cases for TCNQ-F$_2$ \cite{Emge1982} and TCNQ-F$_4$ \cite{Mahns2014}. For CT salts based on TCNQ, the C$\equiv$N stretching modes are dominant in the spectra and the shifting from pure acceptor to CT complex is prominent. 
Besides the shift in C$\equiv$N vibration mode, also the C=C stretching mode of the acceptor is a good indication for the amount of charge transfer \cite{Mene1986}. 
The C$\equiv$N stretching mode is easier to analyze, but as the cyano groups are situated at the end of the acceptor molecule and can be influenced by surrounding molecules, this stretching reflects the CT not as accurate as the C=C vibration in the center of the molecule. In Figure \ref{fgr:IRCN}, we present IR spectra of CT crystals and compare these with spectra on powder of the pure acceptor compounds in the C$\equiv$N stretching mode region. 

For the neutral acceptors, we observed a strong C$\equiv$N stretching mode at 2222.7$\,$cm$^{-1}$ for TCNQ, at 2229.5$\,$cm$^{-1}$ for TCNQ-F$_2$ and at 2227.7$\,$cm$^{-1}$ for TCNQ-F$_4$ (orange arrows in Figure \ref{fgr:IRCN}). According to 
Ref.~\citenum{Bozio1978} these bands can be assigned to b$_{3g}\nu_{42}$ for TCNQ and to b$_{1u}\nu_{18}$ in TCNQ-F$_4$.

Compared to the pure acceptors, the charge transfer compounds show no additional peaks, besides the picene/TCNQ-F$_4$ complex. The relative intensities of the second peak around 2214$\,$cm$^{-1}$ increases compared to the large peak around 2227.7$\,$cm$^{-1}$ as described in Mahns et al. \cite{Mahns2014}. All other compounds show no changes in relative peak intensities. The shift of the C$\equiv$N stretching mode is prominent in all measurements, despite for phenanthrene/TCNQ-F$_4$ and are listed in Table \ref{tbl:IR}. The degree of charge transfer of the shift in the C$\equiv$N stretching mode in infrared absorption spectroscopy can be calculated from the empirical relation \cite{Robles1999}:

\begin{equation}
    \rho_{CN}^{IR} = \frac{2\Delta \nu}{\nu_0(1-\frac{\nu_1^2}{\nu_0^2})}
\end{equation}

Where $\Delta \nu = \nu_0 - \nu_{CT}$  and the subscripts 0, 1 and CT denote the C$\equiv$N stretching modes of the pristine acceptor, the acceptor anion and the charge transfer salt.
A second method to detect the charge transfer is a linear approach between neutral acceptor and fully ionized acceptor with charge of 1$\,$e$^-$ comparing the wave numbers of the respective maximum. The wave number of the C$\equiv$N vibration for the fully ionized acceptor in alkaline salts is 2186$\,$cm$^{-1}$ for TCNQ$^{-1}$, 2188$\,$cm$^{-1}$ for TCNQ-F$_2$$^{-1}$ and  2190$\,$cm$^{-1}$ for TCNQ-F$_4$$^{-1}$ \cite{Bozio1978,Takahashi2014,Mene1986}.

\begin{figure}[h]
\centering
  \includegraphics[width=5cm]{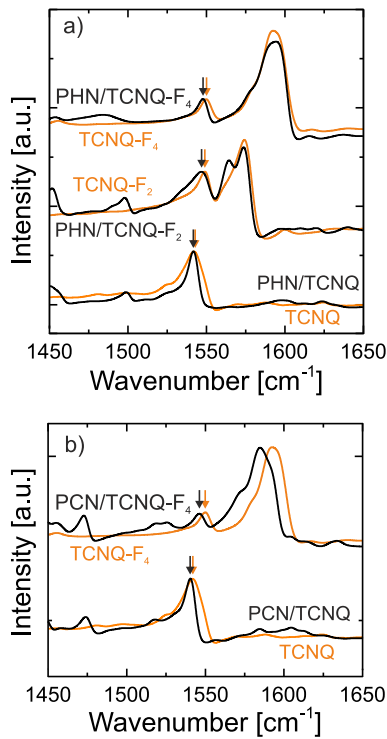}
  \caption{IR spectra in the region of different carbon-carbon stretching modes. The C=C stretching mode outside the ring of the acceptor (orange arrows) is located around 1550$\,$cm$^{-1}$ (b$_{1u}\nu_{20}$ for TCNQ and b$_{1u}\nu_{19}$ for TCNQ-F$_4$). The shift of C=C stretching mode is for a) phenanthrene/TCNQ-F$_x$ but also for b) picene/TCNQ-F$_x$ rather weak, which means a small amount of charge transfer.}
 \label{fgr:IRCC}
\end{figure}

With linear approach, we receive a charge transfer of 0.22$\,$e$^-$ for PHN/TCNQ, 0.28$\,$e$^-$ for PHN/TCNQ-F$_2$ and 0.05$\,$e$^-$ for PHN/TCNQ-F$_4$ when analyzing the C$\equiv$N vibration. This is also in good agreement to results from eq. 2. For the charge transfer salts based on picene, we obtain 0.20$\,$e$^-$ for PCN/TCNQ and 0.16$\,$e$^-$ for PCN/TCNQ-F$_4$, which matches nicely the value of 0.14$\,$e$^-$ in Ref.~\citenum{Mahns2014}. Especially, the results for phenanthrene/TCNQ-F$_x$ are of interest as the charge transfer first increases in the stronger acceptor TCNQ-F$_2$ compared to TCNQ, but decreases again for the strongest acceptor TCNQ-F$_4$. The conclusion from both sets of materials is, that the value of CT is lowest for the strongest acceptor, which is an unexpected result. 
This can be due to the shift in crystal structure and the slightly different surrounding of the C$\equiv$N groups in the structure. As the C$\equiv$N groups are located at the very end of the acceptor molecules, the shift of the IR mode is also strongly depending to neighboring molecules. On the other side, this can be also an indication of a different localization of the charge on the acceptor molecule depending on its fluorination. Therefore, the C=C stretching mode outside the carbon ring in the acceptor needs to be investigated. This shift is more independent to changes in the surrounding and sensitive to another part of the acceptor molecule. These spectra are shown in Figure \ref{fgr:IRCC}. To estimate the CT, we used the linear approach, as discussed before. Therefore, the b$_{1u}\nu_{20}$ mode, identified as the C=C stretching mode, was observed at 1542$\,$cm$^{-1}$ in neutral TCNQ, at 1549$\,$cm$^{-1}$ in neutral TCNQ-F$_2$ and the b$_{1u}\nu_{19}$ mode at 1550$\,$cm$^{-1}$ in neutral TCNQ-F$_4$ and at 1504$\,$cm$^{-1}$ in the TCNQ anion \cite{Bozio1978} with total charge transfer. For the TCNQ-F$_4$ anion the b$_{1u}\nu_{19}$ mode is assigned to C=C stretching and is located at 1500$\,$cm$^{-1}$ \cite{Mene1986}. As TCNQ-F$_2$ is in between TCNQ and TCNQ-F$_4$, we expect the shift of the C=C mode in the anion to be around 1502$\,$cm$^{-1}$. 

 \begin{table}
 \centering
\small
  \caption{IR-frequencies of C$\equiv$N and C=C stretching modes of different charge transfer (CT) salts and pristine acceptors (A) presented in the first two columns. In the third column the difference between these two values $\Delta\nu$ is shown. In the last two columns the charge transfer resulting from linear approach is shown for both the C$\equiv$N and C=C stretchings.}
  \label{tbl:IR}
  \begin{tabular}{llllll}
    \hline
          &   $\nu_{CT} [cm^{-1}]$  &  $\nu_{A} [cm^{-1}]$  &  $\Delta\nu$ &  $\rho_{CT}^{CN} [e^{-}]$ & $\rho_{CT}^{CC} [e^{-}]$\\
    \hline
    PHN/TCNQ  &  2214.6 & 2222.7 &  8.1 & 0.22  & \\
                         & 1541.5 & 1542  & 0.5 &  & 0.01 \\
    PHN/TCNQ-F$_2$ &  2218 &  2229.5  & 11.5 & 0.28 & \\
                                   &  1547 &  1549 & 2.0 &  & 0.04\\
    PHN/TCNQ-F$_4$ & 2225.8 & 2227.7  & 1.9 & 0.05 & \\
                                   & 1547.9 &  1550 & 2.1 &  &  0.04\\
    \hline                               
    PCN/TCNQ    & 2215.4 &  2222.7 & 7.3 & 0.20 &\\
                          & 1540.5 & 1542 & 1.5 &  & 0.04\\
    PCN/TCNQ-F$_4$ & 2221.8 & 2227.7  & 5.9 & 0.16 &\\
                                   & 1546.2 & 1550  & 3.8 &  & 0.08\\
    \hline
  \end{tabular}
\end{table}

With this assumption, we yield a charge transfer of 0.01$\,$e$^-$ for PHN/TCNQ, 0.04$\,$e$^-$ for PHN/TCNQ-F$_2$ and for PHN/TCNQ-F$_4$. For the two picene compounds, we determined a charge transfer of 0.04$\,$e$^-$ for PCN/TCNQ and 0.08$\,$e$^-$ for PCN/TCNQ-F$_4$. The results for CT strongly differ for C$\equiv$N and C=C stretching. For the C$\equiv$N vibration modes, the CT decreases for stronger acceptors (especially for PHN/TCNQ-F$_4$), whereas it increases for the C=C vibration modes with increasing acceptor strength. As described before, the C$\equiv$N vibration modes are located at the end of the acceptor and are probably influenced by their surrounding. In addition, the difference between $\rho_{CT}^{CN}$ and $\rho_{CT}^{CC}$ can be due to an inhomogeneous distribution of the CT over the acceptor molecule. For increasing acceptor strength the contribution of CT might be shifted to the center ring. Therefore, the tendency for the C=C stretching seems more reasonable with a larger amount of CT for increasing acceptor strength although the observed values are probably underestimated. The value of charge transfer lies in all new compounds in the neutral region after definition in \cite{Torrance}. Additional investigation of the electronic structure of these novel complexes using high sensitive spectroscopic measurements are now in progress to investigate the distribution of CT on the acceptor molecule in more detail.

\section{Conclusion}
Single crystals of four new charge transfer complexes with phenanthrene and picene as $\pi$-conjugated donors were obtained by physical vapor transport method. Differential thermal analysis measurements detect the melting temperatures of the respective starting molecules, as well as for PHN/TCNQ. Pressure dependent growth studies improved the crystal growth of these complexes resulting in  thicker crystals sufficient for single-crystal x-ray diffraction. The new crystal structures expand the existing data of different acceptor strengths in the phenanthrene and picene based salts and can be used as model systems to investigate the influence of different acceptors strengths to the amount of charge transfer. Based on these examples and the improvement of the physical vapor transport  technique, further novel materials can be designed. 

\section{Acknowledgement}

The authors acknowledge financial support from of the DFG within the SFB/TR49, the MainCampus program of the Stiftung Polytechnische Gesellschaft Frankfurt am Main (A.M.), the Graduate School of Excellence Materials Science in Mainz and by Centre for Complex Materials COMATT in Mainz (A.C.). We thank F. Ritter, C. Klein and M. Baumgarten for fruitful discussion and K.-D. Luther for technical support.

\section{Supporting Information}
Detailed crystal structure data for each new CT compound are shown in the supporting information.
CCDC files 1448360 (phenanthrene/TCNQ-F$_2$), CCDC 1448361 (phenanthrene/TCNQ-F$_4$), CCDC 1448362 (picene/TCNQ), and CCDC 1448363 (picene/TCNQ-F$_2$) contain the supplementary crystallographic data for this paper and can be obtained free of charge from the Cambridge Crystallographic Data Centre via www.ccdc.cam.ac.uk/data\_request/cif.

\section*{References}

\bibliography{CTbib}

\end{document}